\documentclass[12pt]{article}
\usepackage{epsfig}
\def\be{\begin{equation}}
\def\ee{\end{equation}}
\def\ba{\begin{eqnarray}}
\def\ea{\end{eqnarray}}
\def\ga{\mathrel{\raise.3ex\hbox{$>$\kern-.75em\lower1ex\hbox{$\sim$}}}}
\def\la{\mathrel{\raise.3ex\hbox{$<$\kern-.75em\lower1ex\hbox{$\sim$}}}}

\begin{document}
\baselineskip=16pt
\begin{titlepage}
\begin{center}

\vspace{0.5cm}
{\Large \bf Brane Inflation From Rotation of D4 Brane }\\
\vspace{10mm} {\large Yun-Song Piao$^{a,c}$, Xinmin Zhang$^{a}$
and
Yuan-Zhong Zhang$^{b,c}$} \\
\vspace{6mm} {\footnotesize{\it \large
 $^a$Institute of High Energy Physics, Chinese
     Academy of Sciences, P.O. Box 918(4), Beijing 100039, China\\
 $^b$CCAST (World Lab.), P.O. Box 8730, Beijing 100080\\
 $^c$Institute of Theoretical Physics, Chinese Academy of Sciences,
      P.O. Box 2735, Beijing 100080, China \footnote{
      Email address: yspiao@itp.ac.cn}\\}}

\vspace*{5mm}
\normalsize
\smallskip
\medskip
\smallskip
\end{center}
\vskip0.6in
\centerline{\large\bf Abstract}
 {In this paper, an inflationary model from the rotation of
 D4-brane is constructed.
We show that for a very wide range of parameter,
this model satisfies the observation and find that regarded as
inflaton, the rotation of branes may be more nature than the
distance between branes. Our model offers a new avenue for brane
inflation.
 }
\vspace*{2mm}
\end{titlepage}

Up to date, inflation\cite{GL} is the most likely scenario that
explains many problems of the standard cosmology, such as the
flatness, the horizon and the monopole problems. Although the
recent observations strongly support for the predictions of
inflation, it remains a successful idea seeking for a convincing
theoretical realization. There exist in the literature numerous
phenomenological potentials, which give the proper inflationary
properties\cite{LL}, but some fine-tunings are generically
required in order to obtain sufficient e-folding number before the
end of inflation and to describe the small amplitude of the
observed CMB temperature fluctuations.

It is natural to look for the realizations of inflation within the
string theory, since it is arguably our only known consistent
description of gravity at the shortest distance and highest energy
\cite{P}. Recently, the brane inflationary scenario has been
proposed \cite{DT}, in which the inflaton is identified with an
interbrane seperation. In the brane/antibrane model \cite{BMN},
the exchanges of closed string modes between branes give the form
of inflaton potential, the inflation ends through the condensation
of open string tachyon,
which corresponds to hybrid inflation model \cite{L}. However, the
inflationary scenario based on brane-antibrane attraction and
subsequent collision and annihilation is proved to hardly occur
\cite{BMN, DSS}, since the interaction between brane and antibrane
is very strong, which make the inflaton potential very steep.
Although the hypercubic compactification was considered, a
fine-tuning to the initial condition is required to obtain enough
inflation \cite{BMN, JST}. However, the situation is immensely
improved when the branes intersect at a small angle in some
compact directions \cite{GRZ}. By choosing sufficiently small
angle between the branes the interaction can be made arbitrarily
weak, which make the inflaton potential very flat. The small angle
corresponds to a small supersymmetry breaking, while the
brane/antibrane system is the extreme supersymmetry breaking case,
in which two branes are with opposite orientation. Thus the
inflation may appear naturally in systems that are not so far from
supersymmetry ones. There are also some brane inflationary
scenario implemented in other ways \cite{A, BMQ, BKLO} based on
the string theory.

In this paper, a system which consists of two D4 branes
intersecting at a very small angle $\theta$ is considered, in
which the distance between two D4 brane is much larger than the
string length.
We regard $\theta$ as inflaton and study the inflation of D4 brane
cosmology.
Since the brane system will minimize its energy and make branes
become parallel, when the compactification radii are fixed, the
changing of $\theta$ is mainly connected with that of the homology
class of each brane, in which the homology class of this system is
still conservative for a well-defined theory. We calculate the
amplitudes of the density perturbation and gravitational waves and
give some constraints on the parameters of model.

We assume, before beginning, that the moduli, such as the
compactification radii and the dilaton, are stabilized by some
unknown mechanism at least during the inflationary period. {\it
i.e.} the evolution of the closed string modes is much slower than
that of the open string modes \footnote{Relaxing this condition
\cite{GR}, the inflation from the closed string moduli was
considered in Ref. \cite{BMQ, BKLO}. In fact in our model the
inclusion of dilaton will make the system driven to the minimum of
the run-away potential. The analysis is much more complicated and
goes beyond the scope of this paper. }.We take, not losing
generalization, that the length of all extra dimensions is equal
and denoted by $R$.

As the branes move in a compact space, the R-R tadpole must be
cancelled for a well-defined theory, this condition is equivalent
to the vanishing of the total homology charge of the system. In
addition, the final state after brane inflation is two parallel
D4-brane far away from each other and does not annihilate, thus
the brane cosmology is still in inflationary phase dominated by
the tension of D4-branes and can not exit from it. Generally, an
extra plane with opposite orientation and tension far away from
this two branes system considered can be put, which is not
involved the dynamical evolution of the brane system driving
inflation, to avoid these problems. The angle of D4-brane wrapping
around a torus is determined by the radii of the torus and the
homology class of the brane compactified in it. For our model the
radii of torus are fixed by hand, thus the rotation of D4-brane is
from the changing of its homology class, which is different from
the inflation model of Ref. \cite{BKLO}. The D4 branes wrapping
one cycle in torus of radius $R$ with the $\{n_i ,m_i\}$
homological class have angles $\theta_i = \arctan {m_i\over n_i}$.
This branes system have a total homological charge $\{(n_1+n_2),
(m_1+m_2)\}$. Although the R-R charge of the two D4-branes system
being considered should be a constant, but this does not mean that
the charge of each D4-brane is unchanged. It is this changing that
is considered in our model. For simplification, we take $m_1, m_2
\ll n_1, n_2$, which corresponds very small angles, and $n_1, n_2$
are constants, thus the changing of homology class of each brane
is mainly from $m_1, m_2$. Focusing on the relative angle $\theta
=|\theta_1-\theta_2 |$ between the branes, no losing
generalization, we take $\theta_1=\theta$ and $\theta_2 =0$, which
corresponds to $m_2 =0$. In the small angle approximation $\theta
\ll 1$, the initial energy density of this system is given by \ba
V_{ini}&=&T_4 R \left(\sqrt{n_1^2 +m_1^2}+\sqrt{n_2^2+m_2^2}\right)\nonumber\\
&\simeq & 2T_4 L\left(1+{n_1\tan^2 \theta\over 2(n_1+n_2) }+{\cal
O}(\theta^4)\right) \ea where $T_4 = {M_s^5 \over (2\pi)^4 g_s}$
is the D4 brane tension and $2L =R(n_1+n_2)$ is about the length
sum of each brane warped around the correspondent cycle. The two
branes system would evolve toward the supersymmetry state of lower
energy, {\it i.e.} $\theta\rightarrow 0$, thus the final state of
this system should be two parallel branes with the same total
homological charge. \ba V_{fin} &=&
2T_4 R\sqrt{\left({n_1+n_2\over 2}\right)^2 +\left({m_1+m_2\over 2}\right)^2}\nonumber\\
&\simeq& 2T_4 L \left(1+{n_1^2 \tan^2 \theta\over
2(n_1+n_2)^2}+{\cal O}(\theta^4)\right) \ea The change of energy
between the initial and final state is \be \Delta V = V_{fin}-
V_{ini} \simeq {T_4 L\over 4}\theta^2 \ee

The interaction potential between the two D4-brane intersecting at
a angles $\theta $ can be computed from the exchange of massless
closed string modes, which dual to the one-loop vacuum amplitude
for the open string \cite{P}. The string perturbation theory
computing the interaction between branes requires that the weak
coupling condition $g_s \ll 1$ must be satisfied. In the
limitation of large interbrane separation $y\gg l_s$ and small
angle $\theta\ll1$, the potential is \cite{GRZ, AJ} \be
V(\theta,y) =-{M_s^2\over 8\pi^3}{\theta^3\over y^2} \ee

Therefore, the total effective potential for the two D4 branes
system can be written as \be V_{eff} = \Delta V+V_{int}\simeq {T_4
L\over 4}\theta^2 -{M_s^2\over 32\pi^3} {\theta^3\over y^2}
\label{veff} \ee In the case of $\theta \ll 1$, $g_s\ll 1$ and
$y\gg l_s$, we have $(y l_s^{-1})^2 \gg \theta g_s (M_s L)^{-1}$.
Thus we see that the first term in the $V(\theta, y)$ is much
larger than the sencond term, which is very important to the
calculation of model parameters during inflation in the following.

With the potential (\ref{veff}), the 4D effective action for the
angle $\theta$ and the distance $y$ between the branes can be
written as \footnote{ A detail calculation from supergravity is
given in Ref. \cite{GRZ}. The dynamical term relevant to the angle
$\theta$ is given by small angle approximations. } \be S
={M_p^2\over 2} \int d^4 x \sqrt{-g}R - {1\over 2}T_4 L\int d^4 x
\sqrt{-g} (\partial_{\mu} y\partial^{\mu} y + L^2
\partial_{\mu}\theta\partial^{\mu}\theta) +\int d^4 x \sqrt{-g}
V_{eff}(\theta, y) \label{s}\ee where $M_p$ is the 4D Planck mass,
which is given by a dimensional reduction $M_p^2= M_s^2(RM_s )^6
g_s^{-2}$. The fields $\theta$ and $y$ can be normalized
canonically as \be \phi=\theta\sqrt{T_4 L^3}~~~~{\rm and}~~~~
\psi=y\sqrt{T_4 L} \ee

Assuming that inflation occurs initially at a small angle and a
large distance between the two D4 branes and the fields
corresponding to the angle $\theta$ and distance $y$ is spatially
homogeneous but time-dependent, following definition of Ref.
\cite{WBMR}, the slow-rolling parameters are given

\be \epsilon_{\theta\theta}\simeq \eta_{\theta\theta} \simeq
32\pi^4 {(M_s R)^6\over (M_s L)^3 g_s} {1\over \theta^2} ~~~~~
\epsilon_{yy}\simeq 16\pi^4 g_s {(M_s R)^2\over (M_s L)^3 }
\left({R\over y}\right)^6 \theta^2 \ee \be \eta_{\theta y} \simeq
16\pi^5 {(M_s R)^3\over (M_s L)^3} \left({R\over y}\right) ~~~~~
\eta_{yy} \simeq -48\pi^5 {(M_s R)^2\over (M_s L)^2} \left({R\over
y}\right)^4 \theta \ee In the double fields inflation, the
involved calculations are very complex \cite{WBMR}. But
fortunately, we find that \be {\eta_{\theta\theta}\over \eta_{yy}}
\simeq -{2\over \pi g_s (M_s L)}\left({y\over l_s}\right)^4 \gg 1
\ee \be {\eta_{\theta y}\over \eta_{y y}}\simeq -{y \over 3\theta
L} ~~~~{\rm and}~~~~ {\epsilon_{yy}\over \eta_{yy}}\simeq -{1\over
3\pi}{\theta g_s \over M_s L}\left({R\over y}\right) \ee and for a
very wide range of parameter, we have $\epsilon_{yy}, \eta_{\theta
y}, |\eta_{yy}| \ll\eta_{\theta\theta}$. Therefore, during
inflation the rolling of $\theta$ is far fast than that of $y$,
and in this case, the double fields inflation in this model can be
simplified to a simple field inflation corresponding to $\theta$.

The e-folding number during inflation is \be N=\int H dt \simeq -
{T_4 L^3 \over M_p^2}
\int_{\theta_{60}}^{\theta_{end}}{V(\theta)\over V^\prime
(\theta)} d\theta , \label{n} \ee where $\theta_{60}$ is the field
value corresponding to $N\simeq 60$ required when the COBE scale
exits the Hubble radius, and $\theta_{end}$ is the field value at
which inflation ends, which is determined by
$\eta_{\theta\theta}\simeq 1$. Thus we obtain $\theta_{60}\sim
10\theta_{end}$.

The amplitude of the density
perturbation is given by \cite{LL}
\be
{\cal P}_s\sim {H^2\over \sqrt{T_4 L^3}\dot \theta}
\sim {10^2 g_s\over (M_s L)(M_s R)^3}
\ee
and the correspondent spectrum index is
\ba
n-1 &\simeq& 2\eta_{\theta\theta} -6\epsilon_{\theta\theta}\nonumber\\
&\sim& 10^2\pi^4 {(M_s R)^6\over (M_s L)^3 g_s} {1\over
\theta_{60}^2} \ea The amplitude of the gravitational waves
produced during inflation is \be {\cal P}_g \sim {H\over M_p} \sim
{(M_s L)^{1\over 2}\over (M_s R)^6}g_s^{3\over 2} \theta_{60} \ee
If takeing $g_s \sim 0.01$, $M_s R\sim 10$ and $M_s L \sim 10^6$,
we obtain $\theta_{60}\sim 10\theta_{end} \sim 0.01$, ${\cal P}_s
\sim 10^{-5}$, $n\sim 0.99$ and ${\cal P}_g\sim 10^{-8}$, which
are a set of consistent value satisfying the CMB observation. In
this case, $M_s\sim 10^{-4}M_p$. We can see $H\sim 10^{-8}M_p\ll
M_s$, which avoids the correction of higher curvature terms in
(\ref{s}) and makes the description of the effective theory
reasonable.

Unlike the inflation model of brane/antibrane or branes inflation
with fixed angle, in which the distance between the branes
corresponding to inflaton shrinks to $\sqrt{\theta}M_s^{-1}$, the
tachyonic mode of the open string stretching between the branes
appears, the inflation ends through the tachyon condensation and a
more stable configuration forms \cite{HN}, under our considering
model, the $\theta$ rolls down towards smaller value during
inflation, when $\eta_{\theta\theta}\simeq 1$, the inflation ends
and $\theta$ oscillates about $\theta =0$, the released energy
reheats the brane cosmology. We can estimate the reheating
temperature $T_{reh}$ by regard $T_{reh}^4$ as the initial
potential energy \be T_{reh}\sim M_s\left({(M_sL) \theta^2\over
64\pi^4 g_s}\right)^{1\over 4} \ee Taking above value, we obtain
$T_{reh}\sim 10^{15}$Gev. Since the distance $y$ almost do not
change during inflation, there is not the open string tachyon
appeared in this model. When $\theta =0$, the interaction between
two brane disappears. The final configuration should be two
parallel D4-brane far away from each other, which move towards
each other at a very small relative velocity produced during
$\theta\neq 0$.

In summary, in the approximation of small angle $\theta$ and large
interbrane separation $y$, the double fields inflation of $\theta$
and $y$ from the system, which consists of two D4 branes
interacting a angle, was studied. We found that the dynamical
evolution of $\theta$ is dominated during inflation {\it i.e.}
when the dynamics of both $\theta$ and $y$ is considered, the
$\theta$ inflation may be more nature. Taking $M_s L\sim 10^6$,
$M_s R\sim 10$ and $g_s\sim 0.01$, we obtain the reasonable
amplitude and tilt of the density perturbation satisfying the CMB
observation and the string scale $M_s\sim 10^{15}$Gev, which
corresponds to the GUT scale. In this model, there is not the open
string tachyon appeared which make inflation end, as in the model
of brane/antibrane or branes with fixed angle. The inflation
naturally ends at $\eta_{\theta\theta}\simeq 1$ and the released
energy reheats the brane cosmology. A rough estimation of
reheating temperature is given $T_{reh}\sim 10^{15}$Gev. In fact,
how the energy is released and whether it radiate into the brane
or bulk remain open, a deeper discuss is worth to be done.
Finally, we would like to mention that Halyo have proposed a
rotating-brane inflation model \cite{H}, but our model is in a
different brane setup and parameter regime. Our results offer a
new avenue for brane inflation.

\textbf{Acknowledgments}

We would like to thank Miao Li, Ren-Jie Zhang for interesting
conversations and comments on our manuscript. We also thank
Qing-Guo Huang for discussions. This project was in part supported
by NNSFC under Grant Nos. 10175070, 10047004 and 19925523 as well
as also by the Ministry of Science and Technology of China under
grant No. NKBRSF G19990754.

\end{document}